\documentstyle[preprint,aps]{revtex}

\begin{document}

\draft

\preprint{MA/UC3M/01/95}

\title{Intentionally disordered superlattices with high dc conductance}

\author{Enrique Diez and Angel S\'anchez }

\address{Escuela Polit\'ecnica Superior,
Universidad Carlos III de Madrid,
C./ Butarque 15, E-28911 Legan\'es, Madrid, Spain}

\author{Francisco Dom\'{\i}nguez-Adame}

\address{Departamento de F\'{\i}sica de Materiales,
Facultad de F\'{\i}sicas, Universidad Complutense,
E-28040 Madrid, Spain}

\maketitle

\begin{abstract}

We study disordered
quantum-well-based semiconductor superlattices where the disorder is
intentional and short-range correlated. Such systems consist of
quantum-wells of two
different thicknesses randomly distributed along the growth
direction, with the additional constraint that wells of one kind always
appears in pairs.  Imperfections due to interface roughness are
considered by allowing the quantum-well thicknesses to fluctuate around
their {\em ideal} values.  As particular examples, we consider wide-gap
(GaAs-Ga$_{1-x}$Al$_{x}$As) and narrow-gap (InAs-GaSb) superlattices.
We show the existence of a band of extended states in perfect correlated
disordered superlattices, giving rise to a strong enhancement of their
finite-temperature dc conductance as compared to usual random ones
whenever the Fermi level matches this band.  This feature is seen to
survive even if interface roughness is taken into account. Our
predictions can be used to demonstrate experimentally that structural
correlations inhibit the localization effects of disorder, even in the
presence of imperfections. This effect might be the basis of new,
filter-like or other specific-purpose electronic devices.

\end{abstract}

\pacs{PACS numbers: 73.20.Jc, 73.20.Dx, 72.20.$-$i, 85.42.$+$m}

%\begin{multicols}{2}

\narrowtext

\section{Introduction}

In the past few years, a considerable amount of work has been devoted to
establish that electron localization may be suppressed and bands of
extended states appear in one-dimensional disordered systems whenever
disorder exhibits spatial correlations
\cite{Flo,Wu,Bovier,Datta,JPA,nos2,nos3,Hilke,rusos,ind}.
This unexpected result arising from purely theoretical research is indeed
important because such effect is likely to be of interest for applications.
Unfortunately, these predictions have never been verified
experimentally, and as a consequence there is still some controversy as
to their relevance, their physical implications on transport properties,
and the fabrication of new devices based on those peculiar properties.
Therefore, it is crucial to find physically realizable systems to
ascertain the existence of this new phenomenon.  Since there has been much
theoretical and experimental work in disordered
semiconductor superlattices (SLs)
related to localization electronic effects \cite{multi}, it seems
reasonable to propose random SLs with some kind of structural
correlation as good candidates to check experimentally the validity of
the above mentioned theoretical results.  The advances achieved in
molecular beam epitaxy (MBE), which allow to fabricate SLs tailored with
the desired conduction- and valence-band profiles, support the
feasibility of our suggestion.  On the other hand, previous results of
us on simple, highly idealized models of narrow-barrier SLs indicate
that the effects of correlated disordered should be clearly visible in
such systems \cite{nos2}.  Nevertheless, the theoretical description of
actual SLs requires a more accurate model than the one we have
previously proposed \cite{nos2}.  In particular, effects such as
finiteness of barrier widths, nonparabolicity effects and fluctuations
due to imperfections originated during growth should be taken into
account.  In this paper we concern ourselves with the analysis of all
those effects and study their influence on the theoretically predicted
set of extended states, aiming to clarify whether those
uncontrolled factors modify or
not transport properties of random SLs with intentional correlated
disorder.

The paper is organized as follows.  In Sec.~II, we present our system
and our analytical results on transport properties (transmission coefficient
and dc conductance) of intentionally disordered SLs. In a first stage we
consider the case of decoupled host bands (one-band model) and later we
extend the results to narrow-gap semiconductors, using a two-band model
describing nonparabolicity effects and coupling of host bands.
Correlated disorder is introduced by taking quantum-wells (QWs) with two
different average thicknesses, placing them at random with the
constraint that one of them always appears in pairs.  The body of the
paper is Sec.~III, where we discuss our results on transmission
coefficient and dc conductance for several temperatures.  We show that
the existence of bands of extended states in these structures reveals
itself through well-defined peaks in the dc-conductance.  In addition,
we also consider imperfect SLs by allowing the QW thickness to fluctuate
around their nominal values
and study how this unintentional randomness affects electron transport.
Results are compared with those obtained in uncorrelated disordered SLs.
Finally, in Sec.~IV, we summarize our results and give a brief
explanation, at the actual experiment level, on how our predictions can
be used to demonstrate experimentally that structural correlations
inhibit the localization effects of disorder, concluding by suggesting
possible applications of the so designed devices.

\section{Model and theory}

\subsection{Electronic structure of one-band SLs}

In the simplest picture, the SL potential derives directly from the
different energies of the conduction- and valence-band edges at the
interfaces.  A single QW consists of a layer of thickness $d_A$ of a
semiconductor A embedded in a semiconductor B. In our model of
disordered SL with no imperfections, we consider that $d_A$ takes at
random only two values, $a$ and $a^{\prime}$.  The thickness of layers B
separating neighbouring QWs is assumed to be the same in the whole SL,
$d_B=b$.  A random dimer QW SL (DQWSL) is constructed by imposing the
additional constraint that QWs of thickness $a^{\prime}$ appear only in
pairs, called hereafter a dimer QW (DQW), as shown in
Fig.~\ref{esquema}.  As already mentioned, we also discuss the case of
{\em actual} SLs, where imperfections during growth appear.  We introduce
excess or defect of monolayers during growth by allowing the
width of the layers of semiconductor A to
fluctuate uniformly around the mean values $a$ and $a^{\prime}$.
Therefore, $d_A=a(1+W\epsilon_n)$ or $d_A=a^{\prime}(1+W\epsilon_n)$,
where $W$ is a positive parameter measuring the maximum
deviation from the mean and
$\epsilon_n$ is chosen according to a uniform probability distribution
$P(\epsilon_n) = 1$ if $|\epsilon_n| < 1/2 $ and zero otherwise.  It is
important to stress that $\{\epsilon_n\}$ is a set of random {\em
uncorrelated} variables even when the lattice is constructed with the
dimer constraint.  Therefore, each QW presents a slightly different
value of its thickness and resonant coupling between electronic states
of neighbouring wells decreases.

We focus now on electron states close to the bandgap with ${\bf
k}_{||}={\bf 0}$ and use the one-band effective-mass framework to
calculate the electron wave functions and allowed energies in wide-gap
SLs. Within this approach, the wave function is written as a product of
a band-edge orbital with a slowly varying envelope-function $F(x)$. The
envelope-function satisfies a Ben Daniel-Duke equation with an
effective-mass $m^*(x)$, $x$ being the coordinate along the growth
direction, as follows
\begin{equation}
\label{Schr}
\left[-\,{\hbar^2\over 2}\,{d\phantom{x}\over dx}{1\over m^*(x)}
{d\phantom{x}\over dx}+\sum_{n}^{} V(x-x_n)\right]\>F(x)=E\>F(x),
\end{equation}
with
$$  V(x-x_n) = \left\{ \begin{array}{ll}
               \Delta E_c,  & \mbox{if $|x-x_n|<b/2$}, \\
                0, & \mbox{otherwise},
                \end{array} \right.
$$
where $\Delta E_c$ is the conduction-band offset defined as $E_{cB} -
E_{cA}$ and $x_n$ denotes the position of the centre of the $n\,$th
barrier.  An explicit dependence of both $E$ and $F(x)$ on quantum
numbers is understood and they will be omitted in the rest of the paper.
The energy is measured from the bottom of the conduction-band in the
semiconductor A ($E_{cA}=0$).  Let us consider states below the barrier
($0<E<\Delta E_c$), which are the most interesting ones to study quantum
confinement effects.  The corresponding envelope-function in the QW
between the barriers centered at $x_n$ and $x_{n+1}$ is
\begin{equation}
F^A_n(x)=p_n^Ae^{i\gamma(x-x_n-b/2)}+q_n^Ae^{-i\gamma(x-x_n-b/2)},
\end{equation}
for $x_n+b/2<x<x_{n+1}-b/2$.  Here $\gamma^2=2m_A^*/\hbar^2$, $m^*_A$
being the effective-mass in the QWs. $p_n^A$ and $q_n^A$ are two
constants to be determined later.  Inside the $n\,$th barrier the
envelope-function can be written
\begin{equation}
F^B_n(x)=p_n^Be^{-\eta x}+q_n^Be^{\eta x},
\end{equation}
for $x_n-b/2<x<x_n+b/2$ and now $\eta^2=2 m_B^*(\Delta E_c -
E)/\hbar^2$, $m_B^*$ being the effective-mass in the barriers. Here
$p_n^B$ and $q_n^B$ are also constants.

Imposing continuity of $F(x)$ and $[m^*(x)]^{-1}dF(x)/dx$ at the
interfaces, we can relate the corresponding envelope-function values at
both sides of the $n\,$th barrier via a $2\times 2$ transfer-matrix
$M(n)$ of the form
\begin{equation}
\left( \begin{array}{c} p_n^A \\ q_n^A \end{array} \right)
= M(n) \left( \begin{array}{c} p_{n-1}^A\\ q_{n-1}^A\end{array} \right)
\equiv\left( \begin{array}{cc} \alpha_n & \beta_n \\
\beta_n^* & \alpha_n^* \end{array} \right)
\left( \begin{array}{c} p_{n-1}^A \\ q_{n-1}^A \end{array} \right),
\end{equation}
where we have defined
\begin{mathletters}
\label{elements}
\begin{eqnarray}
\alpha_n&=&\left[\cosh(\eta b)+{i\over 2}\,
\left({\gamma m_B^*\over\eta m_A^*}-{\eta m_A^*\over\gamma m_B^*}
\right)\sinh(\eta b)\right]\,e^{i\gamma(\Delta x_n-b)},
                                  \label{elementsa}\\
\beta_n&=&-{i\over 2}\,\left({\gamma m_B^*\over\eta m_A^*}+
{\eta m_A^*\over\gamma m_B^*}\right)\sinh(\eta b)e^{-i\gamma(\Delta
x_n-b)},                          \label{elementsb}
\end{eqnarray}
\end{mathletters}
with $\Delta x_n\equiv x_n-x_{n-1}$, and $\alpha_n^*$ and $\beta_n^*$
are the complex conjugates of $\alpha_n$ and $\beta_n$ respectively.
Letting $N$ be the total number of barriers, the transfer-matrix $T(N)$
of the SL is obtained as the product
\begin{equation}
T(N) = M(N) M(N-1) \cdots M(1) \equiv \left( \begin{array}{cc}
A_N & B_N \\ B_N^{*} & A_N^{*} \end{array} \right).
\end{equation}
The element $A_N$ can be easily calculated recursively from the
relationship \cite{JPA}
\begin{equation}
A_n=\left( \alpha_n + \alpha_{n-1}^*{\beta_n \over \beta_{n-1}} \right)
A_{n-1} -\,\left( {\beta_n \over \beta_{n-1}} \right) A_{n-2},
\label{A}
\end{equation}
supplemented by the initial conditions $A_0=1$, $A_1=\alpha_1$. The
knowledge of $A_N$ enables us to obtain relevant quantities like the
transmission coefficient at a given energy E, $\tau(E) = 1/|A_N|^2$.
Notice that these expressions are valid for any arbitrary value of QWs
thicknesses and, consequently, they can be used in perfect as well as in
imperfect disordered SLs within the one-band framework.

Finally, once we have computed the transmission coefficient, the
dimensionless finite-temperature dc conductance can be obtained through
the following expression, earlier discussed in detail by Engquist and
Anderson \cite{Ander}
\begin{equation}
\label{kt}
\kappa(T,\mu)= {\int \left(-{\partial f\over \partial E}
\right) \tau(E)dE \over \int \left(-{\partial f\over \partial E}
\right) [1-\tau(E)]}dE,
\end{equation}
where integrations are extended over the allowed bands, $f$ is the
Fermi-Dirac distribution and $\mu$ denotes the chemical potential of the
sample.

\subsection{Electronic structure of two-band SLs}

In this Section we extend our treatment to the case of disordered SLs
made of narrow-gap semiconductors.  Narrow-gap SLs requires a more
complex analysis than those of wide-gap ones.  The scalar equation
arising from the effective-mass approximation is no longer valid at all.
In particular, nonparabolicity effects (an energy-dependent
effective-mass) must be taken into account.  The simplest model that
includes nonparabolicity effects and coupling of host bands is a
two-band Hamiltonian obtained from the ${\bf k}\cdot{\bf p}$ theory
\cite{Beresford,SST}.  The envelope-functions in the conduction- ($F_c$)
and valence-band ($F_v$) satisfy the following Dirac-like equation,
\begin{equation}
\label{Dirac}
\left[ \begin{array}{cc} E_g(x)/2&-i\hbar v \partial \\
i\hbar v \partial & -E_g(x)/2 \end{array}\right]
\left[ \begin{array}{c} F_c(x) \\ F_v(x) \end{array}\right]
=[E-V_g(x)]\left[\begin{array}{c}F_c(x)\\ F_v(x)\end{array} \right],
\end{equation}
where $\partial = d/dx$.  Here $E_g(x)$ denotes the gap ($E_{gA}$ or
$E_{gB}$) of each layer. $V_g(x)$ gives the absolute energy of the
center of the gap and we fix the origin of energies such that it
vanishes in layer $A$.  The parameter $v$, having dimensions of
velocity, is related to the Kane's matrix element and we will consider
it as a constant in the whole superlattice.  This assumption is valid in
most direct gap III-V semiconductors due to the similarities of the
Brillouin zone centre.  Again we appeal to the
transfer-matrix technique to compute the transmission coefficient.  To
this end, let us write down explicitly the solution of (\ref{Dirac}) as
follows
\begin{mathletters}
\label{solution}
\begin{equation}
\left( \begin{array}{c} F_{cn}^A (x)\\ F_{vn}^A(x)\end{array} \right)
=p_n^A \left( \begin{array}{c} 1\\ \rho \end{array} \right)
e^{ik(x-x_n-b/2)}+q_n^A\left(\begin{array}{c}1\\
-\rho\end{array}\right) e^{-ik(x-x_n-b/2)},
\end{equation}
for $x_n+b/2<x<x_{n+1}-b/2$ and
\begin{equation}
\left( \begin{array}{c} F_{cn}^B (x)\\ F_{vn}^B(x)\end{array} \right)
= p_n^B\left(\begin{array}{c}1\\ i\lambda\end{array}\right)e^{-qx}+
q^B_n\left(\begin{array}{c}1\\ -i\lambda\end{array}\right)e^{qx},
\end{equation}
\end{mathletters}
for $x_n-b/2<x<x_n+b/2$. For brevity, we have defined the following real
parameters
\begin{mathletters}
\label{parameters}
\begin{eqnarray}
k      & = & \left( {1\over \hbar v}\right) \sqrt{E^2-E_{gA}^2/4},
               \label{parametersa} \\
\rho   & = & {E-E_{gA}/2\over \hbar v k },
               \label{parametersb} \\
q      & = & \left( {1\over \hbar v}\right) \sqrt{E_{gB}^2/4-(E-V_B)^2},
               \label{parametersc} \\
\lambda & = & {E_{gB}/2-E+V_B\over \hbar v q},
               \label{parametersd}
\end{eqnarray}
\end{mathletters}
where $V_B$ is the energy of the gap centre in the layer B.

Assuming the continuity of the envelope-functions at the interfaces, we
obtain a $2\times 2$ transfer-matrix $M(n)$ whose elements are now given
by
\begin{mathletters}
\label{elementsd}
\begin{eqnarray}
\alpha_n&=&\left[\cosh(qb)+{i\over 2}\,\left( {\rho\over\lambda}-
{\lambda\over\rho} \right)\sinh(qb)\right]e^{ik(\Delta x_n-b)},
              \label{elementsad} \\
\beta_n&=&-{i\over 2}\,\left({\rho\over\lambda}+{\lambda\over\rho}\right)
\sinh(qb)e^{-ik(\Delta x_n-b)}
              \label{elementsbd}.
\end{eqnarray}
\end{mathletters}
By means of the relationships (\ref{A}) and (\ref{kt}) we can calculate
recursively the transmission coefficient along the SL for the two-band
model and then the dc conductance.

\subsection{Transmission through a single DQW}

We now consider a single DQW as shown in Fig.~\ref{esquema}, with the
{\em k\/}th barrier in between, in an otherwise perfect and periodic SL.
We are going to show that there is an specific energy value for which
the so built SL is perfectly transparent.  To this end, we first
consider the condition for energy values to be in an allowed miniband of
the periodic SL, which reads in the one-band model as follows
\begin{equation}
\left|\cos(\gamma a)\cosh(\eta b)-{1\over 2}\,\left({\gamma m_B^*\over
\eta m_A^*}-{\eta m_A^*\over\gamma m_B^*} \right)
\sin(\gamma a)\sinh(\eta b)\right|\leq 1.
\label{condicion1}
\end{equation}
The second condition we have to take into account is simply
Eq.~(\ref{A}) for $n=k, k+1, k+2$; renaming for simplicity
$\alpha_k=\alpha_{k+1}\equiv \alpha'$ and $\alpha_n \equiv \alpha$
($n\neq k, k+1$), eliminating $A_k$ and $A_{k+1}$, and further setting
$\mbox{\rm Re}(\alpha')=0$ we obtain after a little algebra
\begin{equation}
-A_{k+2}=(\alpha+\alpha^*)A_{k-1} -A_{k-2}.
\label{AA}
\end{equation}
Besides a constant phase factor of $\pi$, which has no effects on the
magnitudes of interest, Eq.~(\ref{AA}) reduces to Eq.~(\ref{A}) for a
periodic SL in which barriers $k$ and $k+1$ have been eliminated
from Eq.\ (\ref{elementsbd})
(note that for a periodic SL $\beta_n=\beta_{n-1}$). This
amounts to say that the reflection coefficient at the DQW vanishes and,
consequently, there exists complete transparency at the resonant energy
$E_r$ satisfying $\mbox{\rm Re}(\alpha')=0$, i.\ e.,
\begin{equation}
\cos(\gamma_ra')\cosh(\eta_rb)-{1\over 2}\,\left({\gamma_r
m_B^*\over\eta_r m_A^*}-{\eta_r m_A^*\over\gamma_r m_B^*}\right)
\sin(\gamma_ra')\sinh(\eta_r b)=0,
\label{condicion2}
\end{equation}
where the subscript $r$ refers to the resonant energy $E_r$.
Interestingly, choosing $a'$ appropriately allows us to locate the
resonant energy $E_r$ within an allowed miniband of the periodic SL,
that is, the resonant energy in the range of energies given by
Eq.~(\ref{condicion1}).  Hence, the position of the resonance for which
perfect transmission exists is fixed simply from the values of the
layers thickness and can be tailored as required by choosing appropriate
parameters during growth.

Finally, since Eq.~(\ref{A}) holds also for the two-band model, a similar
resonance condition can be obtained straightforwardly in narrow-gap SLs
using the same arguments as before.  The result is
\begin{equation}
\cos(k_r a)\cosh(q_rb)-{1\over 2}\,\left({\rho_r\over\lambda_r}-
{\lambda_r\over\rho_r}\right)\sin(k_ra')\sinh(q_rb)=0.
\label{condicion22}
\end{equation}
Hence, since there exists also a resonance energy for which the
reflection coefficient at a single DQW vanishes in the two-band
framework, we can conclude that nonparabolicity effects do not suppress
resonant tunneling in the DQW.

\section{Transport through a DWQSL}

The above result concerning resonant tunneling through a single DQW in
an otherwise periodic SL without imperfections does not imply that such
a resonant phenomenon will survive in a disordered SL, that is, when
more than one DQWs are randomly placed in the SL. The transfer-matrix
formalism allows us to compute exactly, although not in a closed
analytical fashion, the transmission coefficient and the dc conductance
at finite temperature in an {\em arbitrary} SL. Thus, in this section we
present the numerical calculation of transport magnitudes in perfect as
well as in imperfect DQWSLs, aiming to elucidate whether resonant
scattering is to be expected in those cases.

As a typical SL described accurately by a one-band model we have chosen
a GaAs-Ga$_{0.65}$Al$_{0.35}$As structure.  In this case the
conduction-band offset is $\Delta E_c=0.25\,$eV, and the effective
masses are $m_A^*=0.067m$ and $m_B^*=0.096m$, $m$ being the
electron mass.  In our computations we have taken $a=b=200\,$\AA\ and
$a^{\prime}=160\,$\AA.  With these parameters we find from
Eq.~(\ref{condicion1}) only one allowed miniband below the barrier,
ranging from $0.116\,$eV up to $0.180\,$eV.  The resonant energy at a
single DQW is $E_r=0.141\,$eV, obtained from Eq.~(\ref{condicion2}), and
thus it lies within the allowed miniband.  To simulate imperfections,
the fluctuation parameter $W$ ranges from $0$ up to $0.05$.  The maximum
value considered here represents excess or defect of three or four
monolayers with the chosen thickness.  This value is above the degree of
perfection now achievable with MBE, so that the results we present are
{\em realistic} in this sense.

As an example of a narrow-gap SL described by the two-band model we
consider nearly lattice-matched InAs-GaSb SL. These two semiconductors
present an almost equal Kane's matrix element leading to $\hbar v =
7.7\,$eV\,\AA, thus supporting our previous assumption that this
parameter is constant through the whole SL. In our case
$E_{gA}=0.36\,$eV, $E_{gB}=0.67\,$eV, and $V=0.665\,$eV, as shown in
Fig.~\ref{fig2}.  We set layer thickness leading to $a=20\,$\AA,
$a^{\prime}=22\,$\AA, and $b=40\,$\AA\ in our numerical computations.
With these parameters we find, as in the one-band system, only one
allowed miniband below the barrier, ranging from $0.607\,$eV up to
$0.719\,$eV.  From Eq.~(\ref{condicion22}) we find that the resonant
energy is now $E_r=0.661\,$eV.

\subsection{Transmission coefficient}

An example of the behaviour of the transmission coefficient $\tau$
around the resonant energy $E_r$ is shown in Fig.~\ref{tran} for a
GaAs-Ga$_{0.65}$Al$_{0.35}$As SL with N=200 barriers.  In
Fig.~\ref{tran}(a) and Fig.~\ref{tran}(b) we show results for perfect
($W=0$) and imperfect ($W=0.05$) DQWSLs, respectively, generated with the
constraint of pairing and with a $1/2$ dimer fraction.  This
fraction is defined as the ratio between the number of wells of width
$a^{\prime}$ and the total number of wells in the lattice.  Since we
have checked that the main conclusions of the present work are
independent of this value, we take a fraction of $1/2$ hereafter.  As a
comparison, Fig.~\ref{tran}(c) shows the transmission coefficient for a
perfect ($W=0$) disordered SL without the constraint of pairing (random
QWSL) with the same number of QWs of thicknesses $a$ and $a^\prime$.  We
observe that for both DQWSLs
close to the resonant energy $E_r$ there is an interval of
energies that shows also very good transmission properties, similar to
that of the resonant energy, in spite of the disordered character of the
SL, even when uncorrelated fluctuations due to imperfections are
present.  On the contrary, this strong enhancement is not observed at all
in random QWSL without pairing. We note that these results are obtained
for a {\em specific} DQWSL; however, we have checked that for different
random realizations of DQWSLs the transmission coefficient behaves
similarly, with only minor changes in its fine structure.

We now discuss the details of the transmission coeffcient behaviour.
We can see an enlarged view of the transmission coefficient for energies
very close to $E_r$ in the insets of Figs.~\ref{tran}(a) and
\ref{tran}(b).  There are several narrow peaks displaying a very high
value of the transmission coefficient.  The number of peaks is related
to the number of wells in the SL; the level spacing would only be
zero in the highly ideal case of an infinite SL\cite{nota}.
In Fig.~\ref{tran}(b) we can see also how large
fluctuations destroy some of those peaks but an important number of them
survive, the smaller the fluctuations the larger this number.
Of course, the location of these peaks is the specific feature of
particular realizations of DQWSLs. From
Fig.~\ref{tran}(c) we conclude that in perfect random QWSLs those peaks
are absent, thus indicating localization of electrons, in contrast to
the situation described in DWQSLs.  Therefore, the loss of quantum
coherence of states close to $E_r$ is much more dramatic in perfect
random QWSLs than in those DQWSLs with relatively large fluctuations,
suggesting that inhibition of localization by structural correlations
is, in fact, a {\em robust} effect.

For brevity we do not show the corresponding results for the two-band
model because the qualitative features are exactly the same.  Thus,
resonant scattering is not destroyed by nonparabolicity effects and
coupling of host bands.  We do discuss them later, in our conductance
study, where we present both results for one and two-band model, so we
postpone any comment to the next subsection.

\subsection{Finite-temperature dc conductance}

So far, we have summarized the main properties of the DQWSL and the
behaviour of transmission coefficient.  One of the main conclusions has
been already mentioned: There is a set of extended states in the DQWSL
in spite of the intentional disorder.  We have provided enough
theoretical evidence and then we can be quite sure of the correctness of
that statement.  The most important point, however, concerns
applications of this result, and this immediately rises one question: Do
these bands of extended states originate {\em experimentally} measurable
features?  We will answer this question in the remainder of the paper.
Specifically, we will devote ourselves to show how the energy of the
resonant states may be determined from finite-temperature dc conductance
measurements.  This would allow us to check whether the predicted energy
close to $E_r$ (recall that $E_r$ depends essentially on the value of
the layer thicknesses and thus it is easily determined, or else the SL
can be built as to show the desired value of $E_r$) agrees with the
measured value in an experimental situation.

We have computed the electrical dc conductance by means of
expression (\ref{kt}) for three different temperatures, $4$, $77$, and
$300\,$K, and for the three kinds of SL that we are studying, namely
DQWSL, DQWSL with fluctuations, and random QWSL.  A global view of the
results is presented in Figs.~\ref{cond1} and \ref{cond2}.  In
Fig.~\ref{cond1} the dc conductance at $77\,$K as a function of the
chemical potential of the sample is seen for the three different SLs,
with the same parameters as in Fig.~\ref{tran}.  A marked peak of finite
width in the dc conductance pattern is clearly observed whenever the
chemical potential lies close to $E_r$ in the perfect DQWSL
[Fig.~\ref{cond1}(a)].  This peak persists when we add imperfections
during growth, even when fluctuations are as large as a $5\%$
[Fig.~\ref{cond1} (b)], and their only appreciable effect is a slight
reduction of its height.  On the contrary, this strong peak is not
observed when DQWs are absent and the SL is purely random
[Fig.~\ref{cond1}(c)].  It is not
difficult to understand why this is so.  The derivative of the
Fermi-Dirac function in Eq.\ (\ref{kt})
at not very high temperatures is very peaked around
the chemical potential.  Therefore, only when the chemical potential
lies close to the set of extended states, that is to say, close to the
resonance, there will be positive contributions to the conductance due
to conducting states.  Of course, if there are no extended states as in
ordinary random QWSLs, never exist positive contributions and the sample
will always show almost zero dc conductance.

In Fig.~\ref{cond2} we present the dc conductance at three different
temperatures for the same DQWSL as in Fig \ref{cond1}.  As temperature
is increased, the derivative of the Fermi-Dirac function broadens and,
consequently, it is not necessary to choose a chemical potential close
to the resonance to obtain a high dc conductance.  In fact, even if it
is placed far from $E_r$ the integrals will include the contribution of
the extended states, leading to an enhancement of the dc conductance.
The peak height decreases because, for higher temperatures and chemical
potentials very close to resonance, not only extended states are
included by the Fermi-Dirac derivative in the integrals, but also a great
number of localized states far from resonance, with low values of
$\tau$, are weighted in the integral.  The behaviour we show in
Figs.~\ref{cond1} and \ref{cond2} coincides then with the intuitive
expectations.

Let us now turn to narrow-gap SLs. In Fig.~\ref{cond3} we present the dc
conductance results at $77\,$K obtained within the two-band framework
for InAs-GaSb SLs, for perfect ($W=0$) DQWSL, imperfect ($W=0.05$) DQWSL
and random ($W=0$) QWSL.  This figure shows that a strong peak of dc
conductance in perfect as well as in imperfect DQWSLs is also observable
in narrow-gap semiconductors, and then the discussions we present for
the one-band model apply to these results as well.  Once again,
extended states appear when correlated disorder exists, producing a
strong enhancement of dc conductance at finite temperature.  This result
confirms our previous statement that nonparabolicity effects and
coupling of host bands do not prevent the existence of extended states.

The extended or localized nature of electronic states close to the Fermi
level can be evaluated from the dependence of the dc conductance on the
number of layers in the SL. The states are extended (localized) when the
dc conductance is constant (decays exponentially) as the SL size
increases, thus leading to an ohmic (nonohmic) behaviour of the sample.
In Fig.~\ref{long} we can see the dependence of the dc conductance at
$77\,$K on the number of barriers for $\mu=E_r$ and for the three
different reference systems (DQWSL without and with imperfections, and
random QWSL without imperfections).  In this case, we present the results
of an average over 100 SLs.
In the perfect DQWSL the behaviour is
purely ohmic, characteristic of extended states.  When fluctuations
are included, a small departure from the perfect ohmic behaviour is
observed, giving rise to an exponential decrease of the dc conductance
as the system size increases, according to the theory of uncorrelated
disordered systems (let us stress once again that fluctuations are
uncorrelated).  Nonohmic behaviour also appears in random QWSLs,
the separation from the ohmic trend being actually dramatic.  Therefore,
even in the presence of fluctuations, electrical conduction is
much higher in
imperfect DQWSLs than in perfect random QWSLs.  It is then quite clear
that this difference would be even larger if fluctuations are to be
taken into account in random QWSLs.

{}From a more theoretical point of view, it is interesting to evaluate
exactly the localization length at the resonant energy.  This can be
done computing the dc conductance at zero temperature, being nothing but
$\kappa_0\equiv\kappa(0,E_r)=\tau(E_r)/[1-\tau(E_r)]$ from
Eq.~(\ref{kt}).  The rate of the exponential decay of this magnitude
versus the number of barriers ---the so called Lyapunov coefficient---
is the inverse of the localization lenght in units of the SL period or,
in other words, the number of QWs over which the wave function spreads.
For brevity we do not show here the corresponding plots since they are
similar to those shown in Fig.~\ref{long}, and we simply quote the main
results.  In GaAs-Ga$_{0.65}$Al$_{0.35}$As DQWSLs we have observed that
the dependence of $\kappa_0$ with the number of barriers $N$ is of the
form $\ln\kappa_0\propto -\gamma(W)N$ where, using a least square fit,
we have obtained that $\gamma(W)=\nu W^2$ ($\nu>0$).  On the contrary,
similar fits in random QWSLs give $\gamma(W)=\nu^{\prime} W^2+\gamma_0$
($\nu^{\prime}>0$ and $\gamma_0>0$).  Therefore, in this case the
behaviour is intrinsically nonohmic even in the limit $W\to 0$.
However, in DQWSLs the parameter $\gamma(W)$ vanishes quadratically for
small values of fluctuations, indicating that the behaviour is almost
ohmic.  Notice that, strictly speaking, the localization length diverges
only at $W=0$ ($\gamma$ vanishes in this limit), and only in this case
electronic states are truly extended.  This agrees with more elaborated
multifractal analysis results \cite{JPA}.  However, the localization
length still remains very large for low level of fluctuations, so that
states are almost unscattered by disorder.  Therefore, they can be
regarded as extended for the SLs with actually available sizes, thus
contributing to electronic transport.

\section{Conclusions}

To summarize, we have studied transport properties at finite temperature
of intentionally disordered SLs with and without DQWs.  We have
demonstrated that there exists a resonant energy for which electronic
states remain unscattered by a single DQW in an otherwise perfect and
periodic SL, due to the resonant coupling between the two neighbouring
QWs forming the dimer.  One of the main points we have found is that
these resonance effects also arise when a finite number of DQWs are
randomly placed in the SL, in spite of the inherent disorder.  Moreover,
this result is independent of the model adopted to describe the SL,
namely one- or two-band Hamiltonians.  Hence we are led to the
conclusion that nonparabolicity effects and coupling of conduction- and
valence-bands do not affect or qualitatively modify this phenomenon.  In
addition, we have demonstrated that those unscattered states reveal
themselves through a dramatic enhancement of the dc conductance at
finite temperature whenever the Fermi level lies close to the resonance,
this effect being more apparent at low temperatures.  Our present
results prove that this enhancement should be experimentally observable
in actual SLs since imperfections inadvertently introduced during growth
do not severely affect the observed increase of the dc conductance, at
least within the available degree of accuracy in MBE techniques.  This
is indeed an important remark from a practical viewpoint since it means
that deviations of few monolayers from the ideal values of the well
thicknesses cannot destroy the quantum coherence required to observe
delocalization and to have dc conductance.
On the contrary, those resonance phenomena are
completely absent in random SLs, even if fluctuations are neglected.  It
is important to stress that the study of delocalization goes beyond the
mere conceptual interest and, actually, new devices may be developed
based in this effect.  For instance, one can choose appropriate layer
thicknesses ($a$, $a^{\prime}$ and $b$) in such a way that $E_r$ lies
close to the Fermi level of the sample, leading to the already mentioned
enhancement of the dc conductance.  In this way it is possible to
disregard all electronic states other than unscattered ones; this may be
the basis of a design of electronic filters. It is also conceivable that
systems whose conductance would change abruptly with temperature could
be fabricated, as for a given chemical potential the conductance of
the sample would jump when the temperature is such that extended states
become involved. In the same way, systems with other peculiar properties
of interest can be thought of.

To conclude, let us also comment that other scattering mechanisms
(phonons, impurities) should also be taken into account in future works
to get insight into these new phenomena.  Nevertheless, on the basis of
this as well as previous works, we believe that they will not modify our
conclusions, inasmuch theoretical calculations on regular SLs where
those effects are neglected describe to a good approximation
actually built SLs.
Finally, a word is in order to draw attention to this problem from
the
experimental viewpoint.  It is clear that there is a fundamental
question pertaining to basic research involved here, namely the
generality of localization phenomena in physical systems.  We have
already discussed this implication in more theoretically oriented works
\cite{JPA}.  Here, we want to insist instead on the fact that
experimental efforts to verify the results we present are required for a
better understanding of delocalization by correlated disorder.  The
necessity of such understanding is clear from the
perspective of technological
applications of SLs. The class of devices we deal with here are but a
first attempt to design microelectronic systems with unexpected
transport properties, its only virtue being their simplicity.  Once the
way is paved to the construction of other devices with exotic properties
by the comprehension
of the relevance of correlations, it is not difficult
to realize that specific-purpose-systems could be built by using more
sophisticated correlation rules.  Such advances will not be possible
unless the simple problem we have been discussing is understood
in actually fabricated devices.

\acknowledgments

It is with great pleasure that we thank collaboration and illuminating
conversations with Fernando Agull\'o-Rueda.  Work at Legan\'es is
supported by the DGICyT (Spain) through project PB92-0248, and by the
European Union Human Capital and Mobility Programme through contract
ERBCHRXCT930413.  Work at Madrid is supported by UCM through project
PR161/93-4811.

\begin{figure}
%\vspace*{0.3 in}
%\setlength{\epsfxsize}{8.8cm}
%\centerline{\mbox{\epsffile{/users/diez/doctorado/two_band/figura1.ps}}}
\caption{Schematic diagram of the conduction-band profile of a SL
containing a DQW.}
\label{esquema}
\end{figure}

\begin{figure}
%\vspace*{0.3 in}
%\setlength{\epsfxsize}{8.8cm}
%\centerline{\mbox{\epsffile{/users/diez/doctorado/two_band/figura1.ps}}}
\caption{Schematic diagram of the conduction- and valence-band profiles
in a InAs-GaSb interface.}
\label{fig2}
\end{figure}

\begin{figure}
%\vspace*{0.3 in}
%\setlength{\epsfxsize}{8.8cm}
%\centerline{\mbox{\epsffile{/users/diez/doctorado/two_band/figura2.ps}}}
\caption{Transmission coefficient $\tau$ versus energy $E$ for (a)
perfect ($W=0$) DQWSL, (b) imperfect ($W=0.05$) DQWSL, and (c) random
($W=0$) QWSL (c).  Every GaAs-Ga$_{0.65}$Al$_{0.35}$As SL consists of
$N=200$ barriers of $b=200\,$\AA\ whereas the thicknesses of QW are
$a=200\,$\AA\ and $a'=160\,$\AA.  Insets of (a) and (b) show enlarged
views of the transmission coefficient around the resonant energy $E_r$.
Note that the scale in figure (c) is much smaller than in the other two
ones.}
\label{tran}
\end{figure}

\begin{figure}
%\vspace*{0.3 in}
%\setlength{\epsfxsize}{8.8cm}
%\centerline{\mbox{\epsffile{/users/diez/doctorado/two_band/figura3.ps}}}
%\vspace*{0.3 in}
\caption{dc conductance at $77\,$K as a function of chemical potential
for (a) perfect ($W=0$) DQWSL, (b) imperfect ($W=0.05$) DQWSL, and (c)
random ($W=0$) QWSL.  The SLs parameters are the same as in
Fig.~\protect{\ref{tran}}.  Note that the scale in figure (c) is much
smaller than in the other two ones.}
\label{cond1}
\end{figure}

\begin{figure}
%\vspace*{0.3 in}
%\setlength{\epsfxsize}{8.8cm}
%\vspace*{0.3 in}
%\centerline{\mbox{\epsffile{/users/diez/doctorado/two_band/figura4.ps}}}
\caption{dc conductance as a function of chemical potential for a DQWSL
with the same parameters as in Fig.~\protect{\ref{tran}} at for (a) $4$,
(b) $77$, and (c) $300\,$K.}
\label{cond2}
\end{figure}

\begin{figure}
%\vspace*{0.3 in}
%\setlength{\epsfxsize}{8.8cm}
%\centerline{\mbox{\epsffile{/users/diez/doctorado/two_band/figura5.ps}}}
%\vspace*{0.3 in}
\caption{dc conductance at $77\,$K as a function of chemical potential
in InAs-GaSb SLs for (a) perfect ($W=0$) DQWSL, (b) imperfect ($W=0.05$)
DQWSL, and (c) random ($W=0$) QWSL.  Every InAs-GaSb SL consist of
$N=200$ barriers of $b=40\,$\AA\ whereas the thicknesses of QW are
$a=20\,$\AA\ and $a'=22\,$\AA.  Note that the scale in figure (c) is
much smaller than in the other two ones.}
\label{cond3}
\end{figure}

\begin{figure}
%\vspace*{0.3 in}
%\setlength{\epsfxsize}{8.8cm}
%\vspace*{0.3 in}
%\centerline{\mbox{\epsffile{/users/diez/doctorado/two_band/figura6.ps}}}
\caption{dc conductance at $77\,$K as a function of the number of
barriers in GaAs/Ga$_{0.65}$Al$_{0.35}$As for $\mu=E_r=0.141\,$eV in
perfect ($W=0$) DQWSLs (upper curve), imperfect ($W=0.05$) DQWSLs
(middle curve), and random ($W=0$) QWSLs (lower curve).  Parameters are
the same as in Fig.~\protect{\ref{tran}}. In this particular plot
we present results of averages over 100 different SLs for each case.}
\label{long}
\end{figure}

%\end{multicols}


\begin{references}

\bibitem{Flo} J.\ C.\ Flores, {\em Transport in models with correlated
diagonal and off-diagonal disorder.}
J.\ Phys.\ Condens.\ Matter {\bf 1}, 8471--8479
(1989).

\bibitem{Wu} P.\ Phillips and H.-L.\ Wu, {\em Localization and its absence:
A new metallic state for conducting polymers.}
Science {\bf 252}, 1805--1812 (1991).

\bibitem{Bovier} A.\ Bovier, {\em Perturbation theory for the random
dimer model.}
J.\ Phys.\ A {\bf 25}, 1021--1029
(1992).

\bibitem{Datta} P.\ K.\ Datta, D.\ Giri, and K.\ Kundu,
{\em Nonscattered states in a random dimer model.} Phys.\ Rev.\ B
{\bf 47}, 10\,727--10\,737 (1993).
{\em Nature of states in a random dimer model:
Bandwidth-scaling analysis.} {\em Ibid.\/} {\bf 48}, 16\,347--16\,356
(1993).

\bibitem{JPA} A.\ S\'anchez and F.\ Dom\'\i nguez-Adame,
{\em Enhanced suppresion of localization in the
continuous Random-Dimer model.} J.\ Phys.\ A
{\bf 27}, 3725--3730 (1994).
A.\ S\'{a}nchez, E.\ Maci\'{a}, and F.\ Dom\'{\i}nguez-Adame,
{\em Suppresion of localization in models
with correlated disorder.} Phys.\
Rev.\ B {\bf 49}, 147--157 (1994); {\em ibid} 15\,428 (Erratum) (1994).

\bibitem{nos2} E.\ Diez, A.\ S\'anchez, and F.\ Dom\'{\i}nguez-Adame,
{\em Absence of localization and large dc conductance in
random superlattices with correlated disorder.}
Phys.\ Rev.\ B {\bf 50}, 14\,359--14\,367 (1994).

\bibitem{nos3} F.\ Dom\'{\i}nguez-Adame, A.\ S\'anchez, and E.\ Diez,
{\em Quasi-ballistic electron transport in random superlattices.}
Phys.\ Rev.\ B {\bf 50}, 17\,736--17\,739 (1994).

\bibitem{Hilke} M.\ Hilke, {\em Local correlation in one- and
two-dimensional discrete systems.} J.\ Phys.\ A: Math.\ Gen.\ {\bf 27},
4773--4782 (1994).

\bibitem{rusos} A.\ S\'anchez, F.\ Dom\'{\i}nguez-Adame,
 G.\ Berman, and F.\ Izrailev,
{\em Understanding delocalization
in the Continuous Random Dimer model.}
Phys.\ Rev.\ B {\bf 51}, in press (1995).

\bibitem{ind} A.\ Chakrabarti, S.\ N.\ Karmakar, and R.\ K.\ Moitra,
{\em On the role of a new type of correlated disorder in extended
 electronic states in the Thue-Morse lattice.}
Phys.\ Rev.\ Lett.\ {\bf 74}, in press (1995).

\bibitem{multi}
M.\ Kasu, T.\ Yamamoto, S.\ Noda, and A.\ Sasaki,
{\em Photoluminescence lifetime of AlAs/GaAs disordered
superlattices.} Appl.\ Phys.\ Lett.\
{\bf 59}, 800--802 (1991).
X.\ Chen and S.\ Xiong,
{\em Optical properties of GaAs/AlAs superlattices with randomly
distributed layer thicknesses.}
Phys.\ Rev.\ B {\bf 47}, 7146--7154 (1993).
A.\ Wakahara, T.\ Hasegawa, K.\ Kuramoto, K.\ V.\ Vong, and A.\ Sasaki,
{\em Photoluminescence properties of Si$_{1-x}$Ge$x$Si disordered
superlattices.}
Appl.\ Phys.\ Lett.\ {\bf 64}, 1850 (1994).

\bibitem{Ander} H.\ L.\ Engquist and P.\ W.\ Anderson,
{\em Definition and measurement of the electrical and thermal
resistances.} Phys.\ Rev.\
B {\bf 24}, 1151--1154 (1981).

\bibitem{Beresford} R.\ Beresford,
{\em Exact eigenfunctions of a two-band semiconductor in a uniform
electric field.}
Semicond.\ Sci.\ Technol.\ {\bf 8},
1957--1965 (1993).

\bibitem{SST} F.\ Dom\'{\i}nguez-Adame and B.\ M\'endez,
{\em Sawtooth superlattices in a two-band semiconductor.}
Semicond.\
Sci.\ Technol.\ {\bf 9}, 1358--1362 (1994).

\bibitem{nota}
We have confirmed this assertion by
using the Poincare-map formalism developed by us in
Ref.~\onlinecite{JPA}, which allows us to study transmission properties
of a disordered SL of finite length embedded in an infinite and periodic
($d_A=a$) SL; we do not dwell further in this matter because this work
is devoted to the study of realistic superlattices which can be built
and measured.
\end{references}
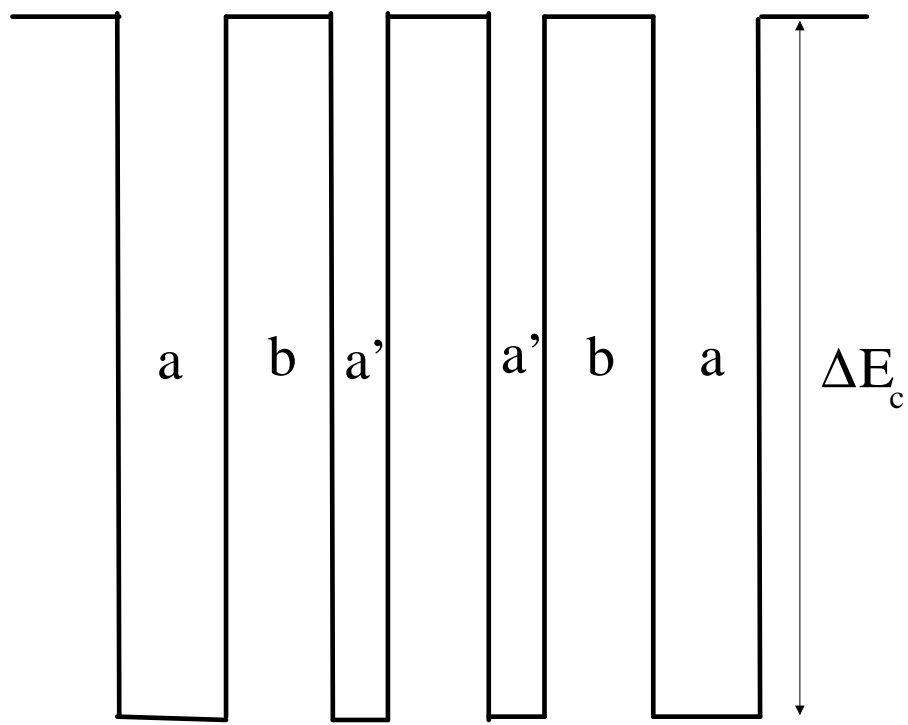
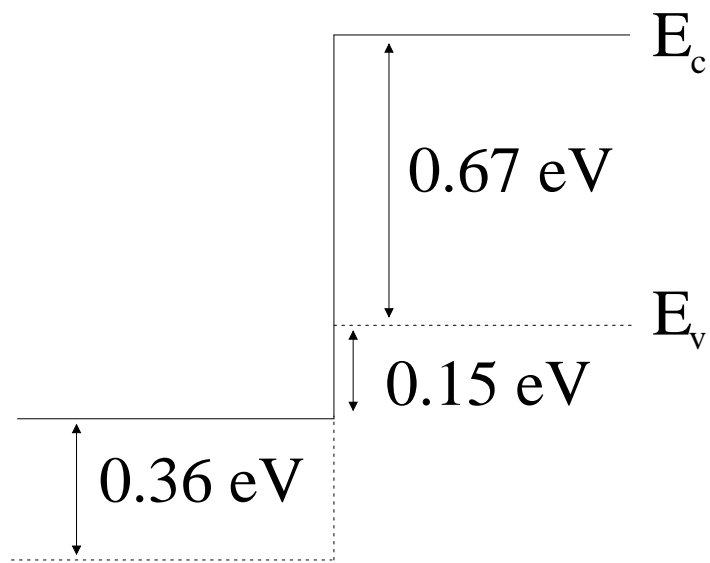
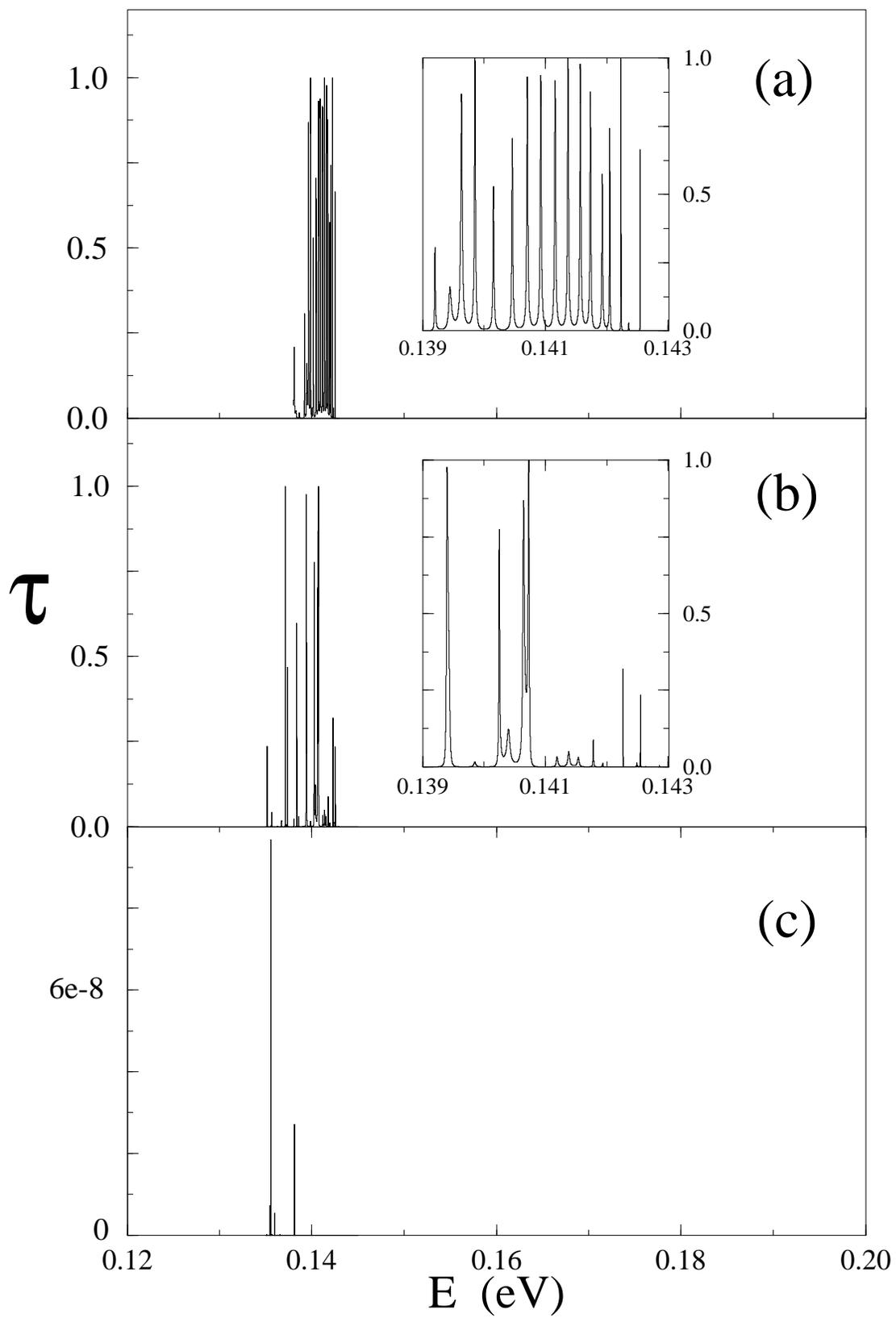
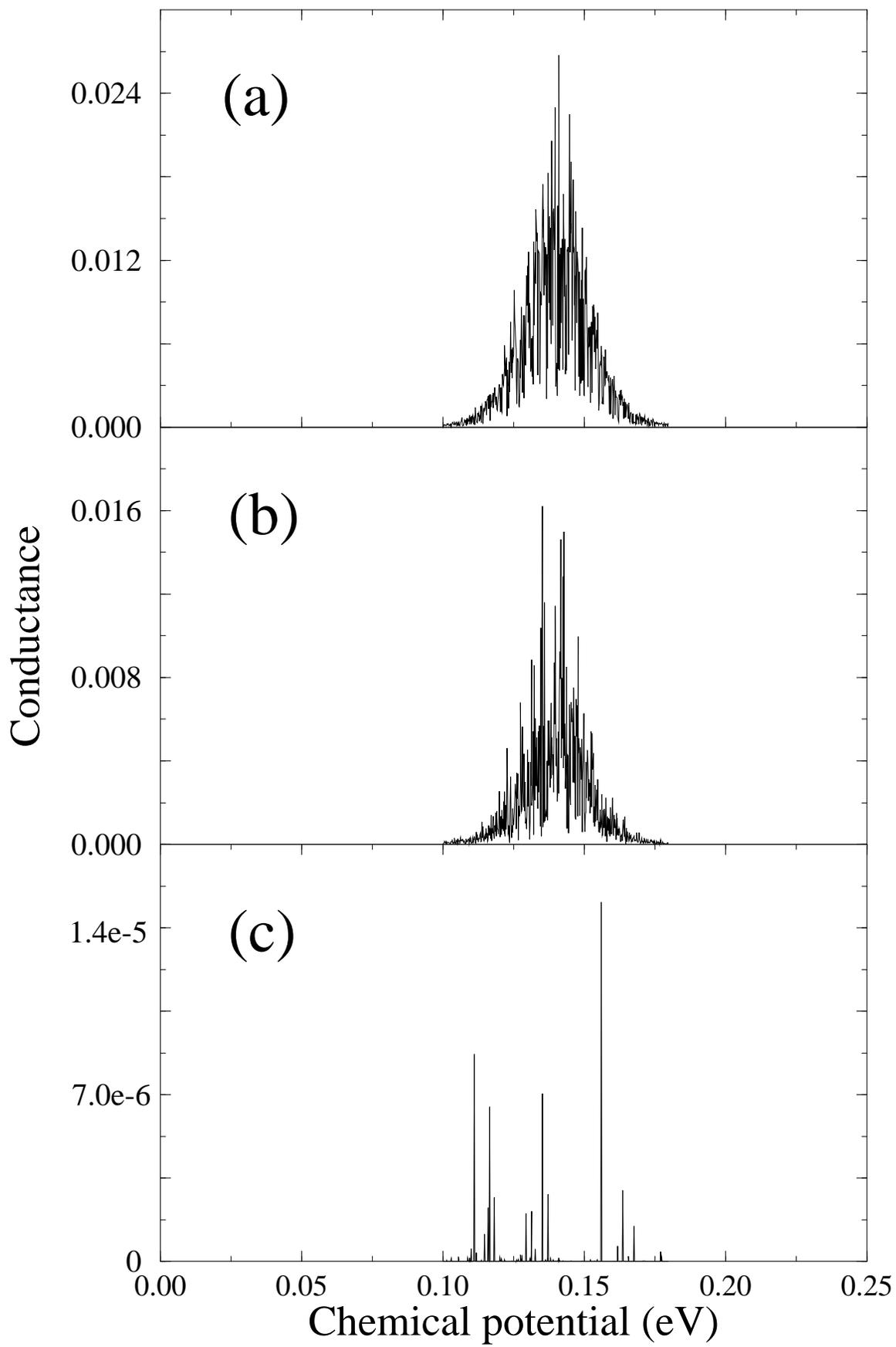
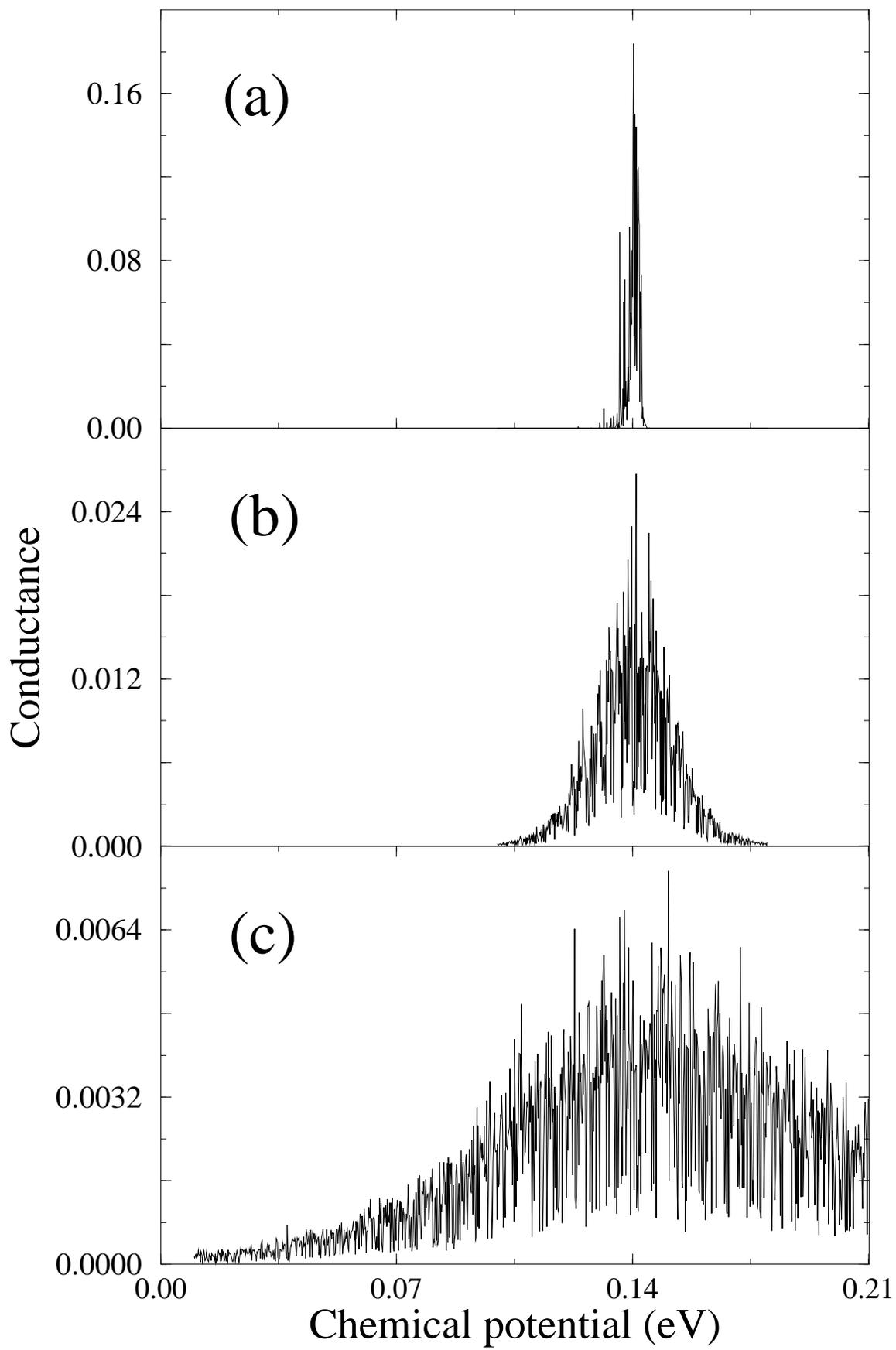
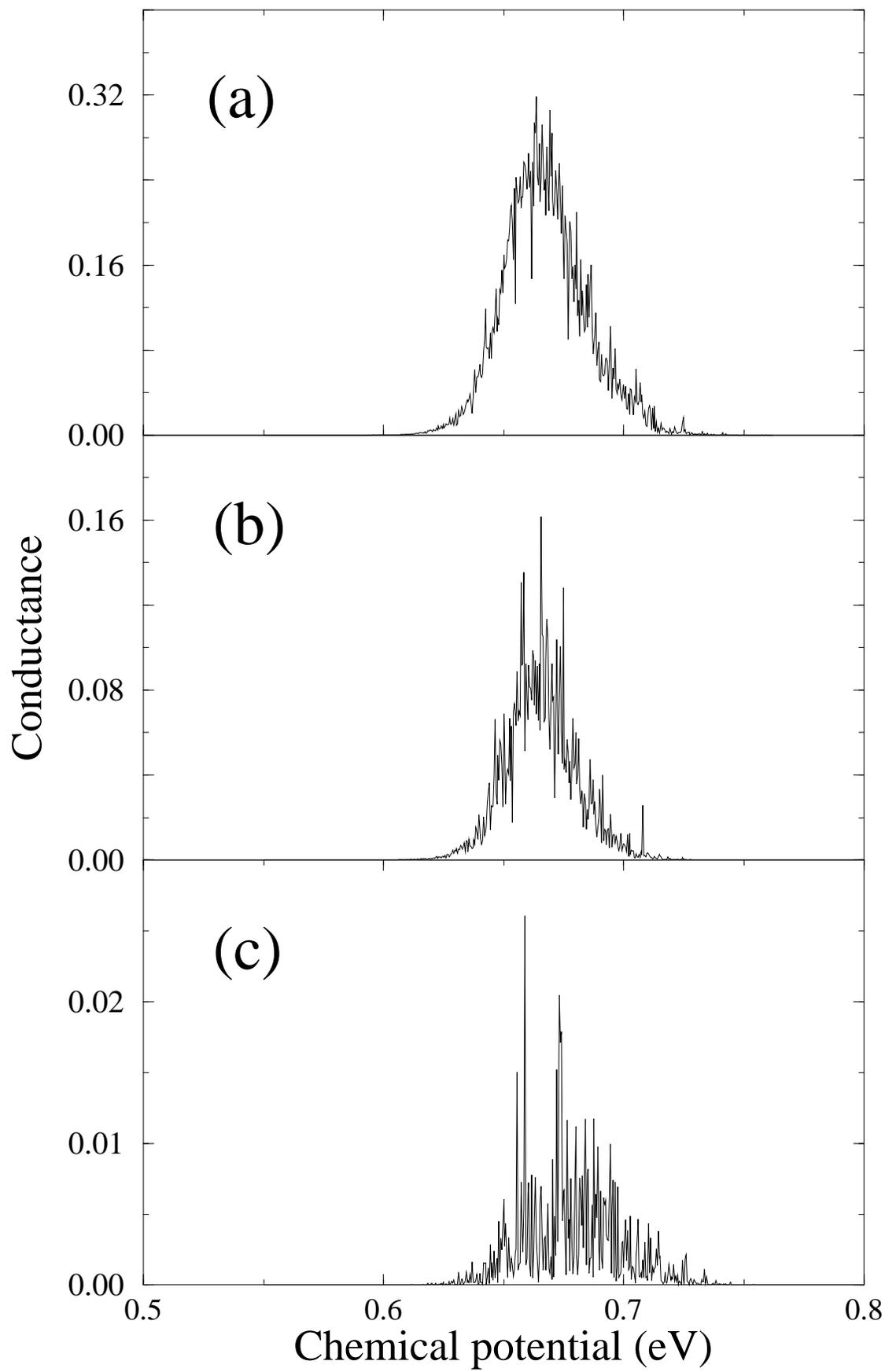
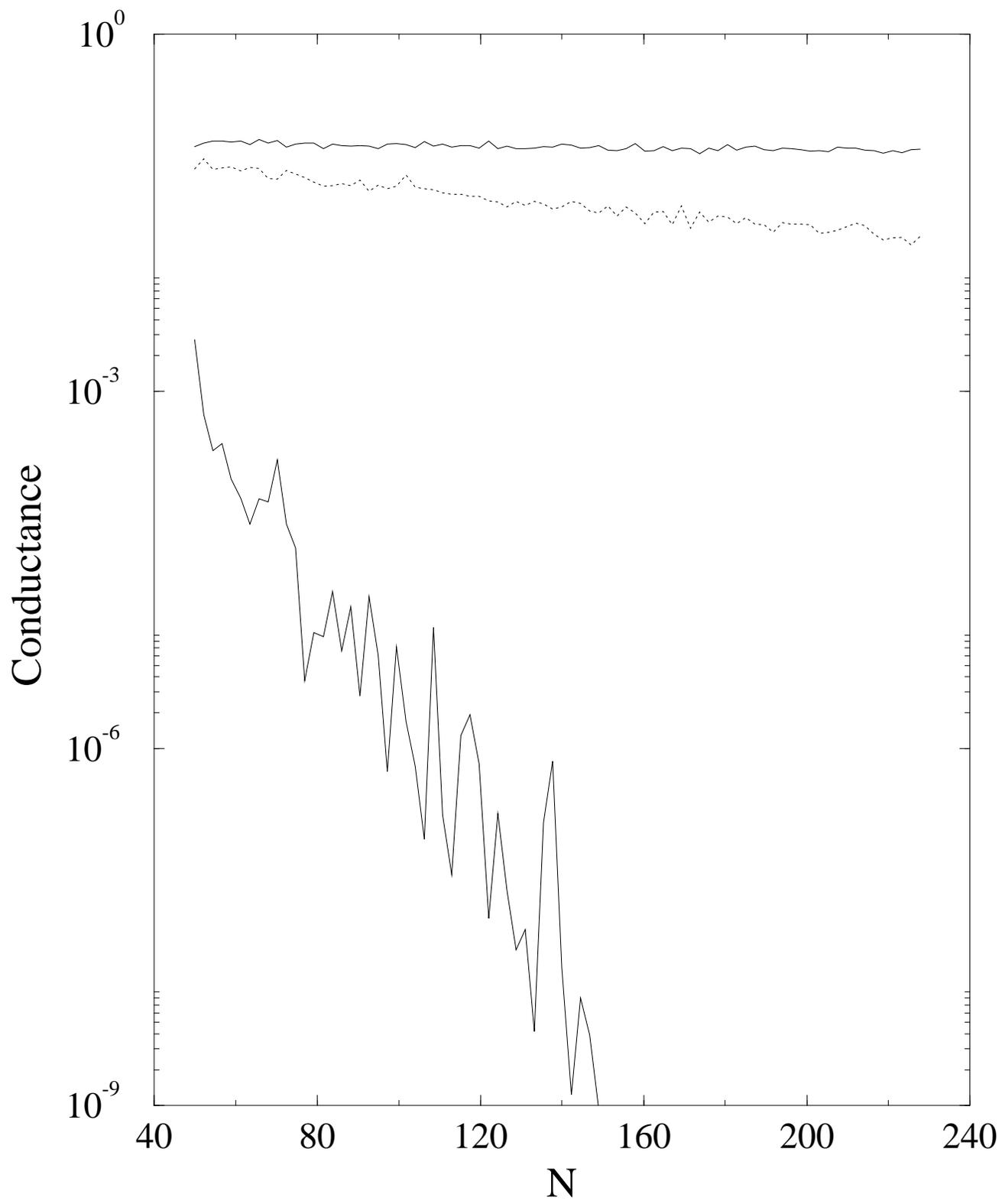
\end{document}